\documentclass[preprint,aps]{revtex4}

\usepackage{graphicx}
\usepackage{epstopdf}
\usepackage{dcolumn}
\usepackage{bm}

\usepackage{color,ulem}
\usepackage{soul}

\usepackage{setspace}

\pdfoutput=1

\begin{document}
\title{Coherent absorption and enhanced photoluminescence in thin layers of nanorods}
\author{G. Pirruccio$^{1}$, G. Lozano$^{1}$, Y. Zhang$^{1}$, S. R. K. Rodriguez$^{1}$, R. Gomes$^{2}$, Z. Hens$^{2}$, and Jaime G\'{o}mez Rivas$^{1,3}$}
\address{$^1$ FOM Institute for Atomic and Molecular Physics AMOLF, c/o Philips Research Laboratories,
        High Tech Campus 4, 5656 AE, Eindhoven, The Netherlands.\\
         $^2$ Physics and Chemistry of Nanostructures, Ghent University, Krijgslaan 281-S3, 9000 Ghent, Belgium.\\
         $^3$ COBRA Research Institute, Eindhoven University of Technology, P.O. Box 513, 5600 MB Eindhoven, The Netherlands.}
\email{pirruccio@amolf.nl}
\date{\today}

\setstretch{2}

\begin{abstract}
We demonstrate a large light absorptance (80\%) in
a nanometric layer of quantum dots-in-rods (QRs) with a thickness of
23 nm. This behavior is explained in terms of the coherent
absorption by interference of the light incident at a certain angle
onto the very thin QR layer. We exploit this coherent light
absorption to enhance the photoluminescent emission from the QRs. Up
to a seven- and five-fold enhancement of the photoluminescence is
observed for \textit{p}- and \textit{s}-polarized incident light, respectively.
\end{abstract}
\maketitle

\section{Introduction}

With the development of nanotechnologies it has become possible to
scale down the dimensions of electro-optical devices. For certain
applications, e.g., in photovoltaics and photodetection, it has
become crucial to find new methods to enhance the optical absorption
in ultra-thin layers of different
materials.~\cite{Stuart1996,Atwater2010} Consequently, enhanced
optical absorption by nanostructures and metamaterials has been
intensively investigated in recent
years.~\cite{Stuart1996,Shaadt05,Panoiu07,Landy2008,Bandiera2008,Driessen09,Aubry2010}
The concept of Coherent Perfect Absorption (CPA) has been recently
introduced.~\cite{Wan11} In particular, it has been shown that is
possible to trap a narrow window of wavelengths of counter
propagating beams inside a thick slab of a given material through
the realization of interference.~\cite{Chong10} CPA is,
thus, described as a combination of interference and dissipation.

In this article we exploit the CPA concept by making use of the
Attenuated Total Reflectance (ATR) technique. Our structure consists
of an ultra-thin layer of absorbing material, with a thickness of 23
nm. We choose colloidal quantum dots in rods (QRs) as the
constituent material of the layer. Quantum dots are a class of materials with
characteristic optical properties, such as the tunability of their
emission spectrum, and their high quantum efficiency if their
surface is well passivated. ~\cite{Reiss09} We have designed the sample and the
experiment such that an efficient conversion of incident light into
pholuminescence is achieved by means of enhanced absorption. In
particular, we demonstrate a significant enhancement of the absorptance of the incident
light at $\lambda$=457 nm onto the layer (80\% for \textit{s}-polarized light
and 70\% for p-polarized light) at a certain angle of incidence,
which leads to a seven- and five-fold enhancement of the QR
photoluminescence for \textit{p}- and \textit{s}-polarized excitation respectively.

ATR is an extended technique used to couple plane waves to
evanescent modes in which a prism is used to enable this coupling. A
typical ATR measurement is characterized by a minimum in the
specular reflectance at an angle of incidence larger than the
critical angle for total internal reflection at the prism-structure
interface. An interpretation for the ATR resonance, based on quantum
interference, has been provided:~\cite{Herminghaus94,Boardman82}
when the momentum-matching condition between the incident wave and
the evanescent mode is realized an additional photon path, besides
the radiative scattering, opens. The relation between this
interpretation and the CPA is provided by the interference nature of
the two phenomena. In the ATR experiment, the interference is due to
the superposition of the probability amplitudes of two events: the
photon scattered out because of the total internal reflection, and
the same photon coupled first into an evanescent mode and coupled
out into radiation. In the resonant condition the radiative damping
of the evanescent mode equals the internal damping due to losses in
the material, and all the light is absorbed. The quantum
interpretation can be rephrased classically in terms of interference
of electromagnetic waves, in which the first photon path corresponds
to the reflected wave at the prism-structure interface and the
second one to the leaky wave associated with the evanescent
mode.~\cite{Raether} This classical interpretation is similar to the
so-called critical coupling by which it is possible to perfectly
transfer energy between two optical media.~\cite{Yariv00,Yariv02}
Similarly to ATR, in the critical coupling theory, also there is a
relation between the internal and the out-coupling losses, namely,
when they are equal it is possible to absorb 100\% of the incident
light.~\cite{Choi01} \newline CPA, ATR and critical coupling are phenomena
characterized by a general scheme, i.e., in all of them optical
interference and dissipation lead to an efficient electromagnetic
energy transfer into a structure and an enhanced absorption. In our
experiments the structure to which we want to transfer the energy is
the QR layer.

\section{Sample preparation and characterization}
We have used CdSe/CdS quantum dots-in-rods as the constituents of
the nanometric layer. The QRs have been synthesized following the
procedure described in Ref. ~\cite{Carbone07}. They have an average
length of 36 nm and an average diameter of 4 nm. The absorbance and
normalized emission spectra of as grown CdSe/CdS nanorods in
suspension in toluene is shown in Fig. 1(a). The large ratio between absorbance at $\lambda$=460 nm and at $\lambda$=611 nm
limits the self-absorption. The quantum efficiencies of the QRs suspended in toluene and
forming a layer in air have been determined to be $70\%$ and $40\%$, respectively. The reduction of photoluminescence quantum efficiency of the rods in the layer is explained by the increased probability for energy transfer between adjacent rods. In particular, an exciton in a close-packed QR layer has a higher probability to end up in a non-emitting rod, which leads to a decrease of the quantum efficiency of the ensemble of nanorods.

Since our goal is to exploit the CPA principle in the ATR
configuration, we have fabricated a sample with the structure
schematically represented in Fig. 1 (b). A suspension of quantum
dots-in-rods in toluene is spin-coated on top of a silica substrate with a refractive index of 1.45.
After baking the sample for 2 minutes at $80{^\circ}$ C, a high-density layer of QRs with a thickness of 23 nm is formed on the substrate. Figure 2 (a) displays a scanning electron microscope (SEM) image of the layer tilted in order to appreciate its thickness. A silica layer with refractive index 1.46 and a thickness
of 350 nm, indicated as matching layer in Fig. 1 (b), is evaporated
on top of the QR layer. The thickness and refractive index of this
layer are critical to achieve coherent absorption in the QR layer.
The optical constants and the thicknesses of the QR and matching
layers have been determined with spectroscopic ellypsometry. Figure
2(b) displays ellypsometry measurements of the real, $n$ (black curve),
and imaginary, $k$ (gray curve), components of the refractive index
of the QR layer. Two peaks are visible at $\lambda$=460 nm and
$\lambda$=600 nm in the curve of the imaginary component of the
refractive index. These peaks correspond to absorption due to
resonant excitation of excitons in CdS and CdSe respectively.

An F2 Schott optical glass prism, with refractive index 1.62 at
$\lambda= 600$ nm, on top of the sample is used in the ATR
configuration. In order to obtain a good optical contact between the
prism and the matching layer and to avoid multiple reflections that
will destroy the interference in the QR layer, we have used a
refractive-index-matching liquid with the same refractive index as
F2.

\section{Reflection and absorption measurements}

The ATR measurements were performed as follows: The prism was
illuminated with a collimated \textit{s}- or \textit{p}-polarized beam from a halogen
lamp, varying the angle of incidence, $\theta$. The detector was
placed at the angle of specular reflection. The specular reflection
was measured in steps of $\theta = 0.1{^\circ}$ by rotating the
sample and a multimode optical fiber coupled to a spectrometer with
a computer controlled rotation stage. In order to obtain the
specular reflectance (\textit{R}) of the QR layer, the reflection
measurements were normalized by the reflection from an angle above
the critical angle for total internal reflection at the
prism-matching layer interface, i.e., an angle at which the reflectance is 1. For
angles larger than the critical angle the transmission through the
sample was negligible.

In Fig. 3 we display the measured (a) and the calculated (b)
specular reflectance (color scale) for \textit{s}-polarized light as a function
of the angle of incidence and the wavelength. The calculation has
been performed using the transfer matrix method for a multilayer
structure.~\cite{Yeh} In these calculations we have fixed the
thickness and refractive index of the QR and matching layer as derived
from the ellipsometry measurements. Therefore, we do
not use any free parameter to fit the measurements. Figures 3 (c) and
3 (d) display cuts of the reflectance measurements (open circles) and
calculations (solid curves) at 611 nm and 457 nm, respectively. These
wavelengths correspond to the wavelength of maximum emission of the
QRs and to the wavelength used to excite the QRs in the
photoluminescence experiments shown in the next section. The strong
increase in the reflectance for angles larger than $\theta \simeq
64{^\circ}$ is due to the total internal reflection at the
prism-matching layer interface above the critical angle
($\theta_{\rm c} = 64.5{^\circ}$ for $\lambda$=700 nm). The
pronounced dip in reflectance around $\theta= 65.5{^\circ}$ and
$\lambda$=450 nm is due to the coherent absorption. The excellent
agreement between measurements and calculations confirms the
validity of the transfer matrix method to describe the light
scattering in the multilayer structure. This remarkable result,
in view of the inhomogeneities in the QR layer [see Fig. 1(b)], can be explained from the fact that ellipsometryc measurements include the effect of the roughness as an effective property of the layer.

For angles larger than the critical angle we estimate the experimental
absorptance in the QR layer as 1-\textit{R}, assuming that the transmittance through the sample in this angular range is negligible. Figure 4(a) displays with open
circles the \textit{s}- (black circles) and \textit{p}-polarized (grey
circles) absorptance for
$\theta$$>$$\theta_{\rm c}$ at $\lambda$=457 nm. The dotted vertical line in this figure
indicates the critical angle. The maximum in absorptance reaches the
extraordinarily large value of 80\% at $\theta = 65.5
{^\circ}$ and 70\% at $\theta=64.2{^\circ}$ for \textit{s}- and \textit{p}-polarized
illumination, respectively. This is a remarkable result considering
that absorption takes place only in the 23nm-thick QR layer. The
shift of the maximum in absorptance between \textit{s}- and \textit{p}-polarized light
can be explained by the different phase shifts in the reflection at
the interfaces for the two polarizations that modifies the interference
condition for coherent absorption. We note that the absorption
length, i.e., the length over which the intensity decreases by a
factor 1/\textit{e} under plane wave illumination, at $\lambda$=457 nm in a thick layer of a material with
the refractive index of the QR layer is on the order of 350 nm for
both polarizations. Therefore, coherent absorption in the QR layer
leads to a reduction of the absorption length by more than 1 order
of magnitude.

In Fig. 4(b), we plot the calculated absorptance at $\lambda$ = 457
nm for \textit{s}-polarization (gray curve) and \textit{p}-polarization (black curve)
using optimized values for the thicknesses of the matching layer to
achieve perfect absorptance at a certain angle. This thickness
corresponds to 480 nm for \textit{p}-polarization and 230 nm for
\textit{s}-polarization. The difference in the two optimum thicknesses is due to the different amplitudes and phases of the fields reflected at each interface of the multilayer structure for \textit{p}- and \textit{s}-polarization. This difference makes it impossible to achieve 100\% absorption for unpolarized light
in a single sample. Nevertheless, the sample used in our experiments,
with a matching layer of 350 nm, exhibits a significant enhancement
of the absorption for both polarizations. We note from the measurements of Fig. 4 (a) and (b) that with the non-optimum thickness of the matching layer we are not truly fulfilling the condition for critical coupling. The maximum absorptance equals 80\% which indicates a 20\% radiation loss in the experiment.

The interference mechanism leading to the coherent absorption in the
QR layer can be appreciated by calculating the field intensity in
the multilayer structure. This calculation is displayed in Fig. 5
for \textit{s}-polarized incident light with a wavelength of 457 nm at four
different angles, namely, $\theta=62{^\circ}$ (a), $64.2{^\circ}$
(b), $65.4{^\circ}$ (c) and $66.5{^\circ}$ (d). The color scale in
these figures represents the spatial distribution of the field
intensity normalized by the incident intensity. The absorbed
electromagnetic power in a medium is given by
$\int_{V}\frac{\omega}{2} \epsilon ''\overline{|E|^2}
dV$,~\cite{Landau84} where $\omega$ is the angular frequency of the
wave, $\epsilon ''$ is the imaginary component of the permittivity
of the medium and \textit{E} is the total electric field where the overbar
indicates the time average over a period. In our system the only
absorbing material with $\epsilon ''\neq0$ is the QR layer and the
integral needs to be evaluated in the volume occupied by this layer.
At $\theta=62{^\circ}$ [Fig. 5(a)] the field intensity in the QR
layer is low. Therefore, for this angle of incidence also the
absorptance in the QR layer is low. At $64.2{^\circ}$ [Fig.
5(b)] the angle of incidence approaches the angle for coherent
absorption. Consequently, the field intensity in the QR layer is
higher. The normalized field intensity is also high
at the prism-matching layer interface due to total internal
reflection. This situation changes drastically at $65.4{^\circ}$
(Fig. 5(c)), where the coherent absorption condition is reached. For
this angle of incidence the fields interfere constructively in the
QR layer where the intensity increases. This field enhancement in
the layer leads also to the enhancement of the absorbed power in the
otherwise weakly absorbing QRs. For this angle of incidence, we
observe a reduction of the intensity at all locations other than the QR
layer. This reduction is the result of the concomitant destructive
interference required by the conservation of energy. At
$\theta=66.5{^\circ}$ the absorptance is low [Fig. 5(d)]. For this angle
the normalized field intensity in the QR layer is also low, while at
the prism-matching layer interface it is maximum due to total internal
reflection.

A description of the mechanism leading to coherent absorption in our
system, related to the aforementioned description of ATR, considers
the coupling of the incident radiation into the quasi-bounded
fundamental mode supported by a lossy waveguide. The waveguide is
defined by the QR layer with an effective index of refraction larger
than the surrounding media and the losses are given by absorption in
the layer and through radiation into the prism. To illustrate this
explanation, we have calculated the eigenfrequencies associated with
the fundamental transverse electric field eigenmode (TE$_{0}$ mode) of a waveguide
with a thickness of 23 nm and a permittivity equal to the effective
permittivity of the QR layer obtained from ellipsometry. The effect of the roughness was taken into account in these calculations since the imaginary component of the effective refractive index, obtained from ellipsometry, includes a reduction in the intensity of the specular reflection due to both absorption and scattering in the QR layer. The
eigenfrequencies are represented as a function of the wavenumber of
the mode in the dispersion diagram shown in Fig. 6 with the solid
line. In these calculations, we consider the layer embedded in a
homogenous dielectric, i.e., we do not consider the prism. Despite
this approximation there is a good agreement with the measured data
represented by the open circles in Fig. 5. These measurements have
been obtained from the attenuated total reflectance data (Fig. 3) by
determining the wavelengths and angles of minimum reflectance and
representing them as a function of the frequency, $\nu=c/\lambda$,
and the in-plane wavevector,
$k_{\parallel}=\frac{2\pi}{\lambda}\sin(\theta)$. The minimum in the reflectance measurements of Fig. 3 (d) can be described as the result of destructive interference between the leakage radiation of the TE$_{0}$ mode through the prism and the specular reflection of the incident beam at the prism-matching layer interface.

\section{Photoluminescence}

Coherent absorption can be exploited to enhance the
photoluminescence from the QR layer. The inset of Fig. 7 (a) shows
a schematic representation of the experimental configuration used
in the photoluminescence experiments. The angle of incidence $\theta$
is varied in these measurements, while the direction of detection,
defined by the angle $\alpha$ between the normal to the layer and
the detector, is kept fixed at $37^\circ$. The QR layer was excited
with a $\lambda=457$ nm laser and incident power of 1.6 mW, using
both \textit{s}- and \textit{p}-polarization. The emission was collected without
polarization selection. From the reflection measurements of Fig.
3(a), we can conclude that the sample does not exhibit any resonance
above 600 nm. The QR photoluminescence at these wavelengths follows
a Lambertian curve as a function of the emission angle. This
Lambertian emission was confirmed in angular dependent emission
experiments (not shown here). In Fig. 7 are shown the spectra taken
for different values of the angle of incidence, $\theta$, and
polarizations. Black curves correspond to \textit{s}-polarized incidence and
gray curves to \textit{p}-polarized incidence. Figure 7 (a) shows the photoluminescence
intensity for $\theta=62^\circ$, i.e., for an illumination angle
below the critical angle. Figures 7 (b) and (c) correspond to the
angles of illumination at which maximum absorption at $\lambda$=457
nm is achieved in the QR layer for \textit{p}- and \textit{s}-polarization
respectively, i.e., $\theta=64.2^\circ$ and $\theta=65.4^\circ$. An
increased emission is achieved at the angles of incidence at which
coherent absorption is observed. Figure 7 (d) shows the
photoluminescence emission due to evanescent excitation of the QR
layer at $\theta=66.5^\circ$.

To directly correlate the absorptance to the photoluminescence
enhancement, we have plotted in Fig. 8 (a) and (b) the absorptance
at $\lambda$=457 nm (solid curves) and the photoluminescence
intensity integrated from 550 nm to 670 nm (squares), as a function
of the angle of incidence for \textit{s}- and \textit{p}- incident polarization,
respectively. The dashed vertical lines correspond to the angles of
Fig. 7 (a-d). As expected, the enhancement of photoluminescence is
directly related to the enhancement of the absorption at the
excitation wavelength of the QRs. It is possible to evaluate an
enhancement factor of the photoluminescence by dividing the value of
the integrated intensity at the angle of maximum emission by the
intensity at an angle lower than the angle of total internal
reflection. Enhancement factors of 4.6 and 6.8 at $\lambda$=457 nm are obtained for \textit{s}-
and \textit{p}- polarization, respectively.

\section{Conclusions}

We have demonstrated experimentally an absorptance up to 80\% for
\textit{s}-polarized light and 70\% for \textit{p}-polarized light at $\lambda=457$ nm
in a 23nm-thick layer of quantum dots-in-rods. This extraordinary
absorption is explained in terms of coherent absorption: The
incident light is trapped in the layer due to the constructive
interference of the fields scattered at different interfaces. The
enhanced absorption is efficiently converted into photoluminescence
from the QR layer. We have obtained a seven-fold enhancement of the
QR emission for \textit{p}-polarized incident light and a five-fold emission
enhancement for \textit{s}-polarized incident light as a result of the
coherent absorption in the layer. This enhancement can be further
increased by optimizing the interference in the multilayer
structure. Enhanced optical absorption in nanometric layers is
relevant for applications in various fields such as photovoltaics
and sensitive photodetection.

\section{Acknowledgement}
This work was supported by the Netherlands Foundation Fundamenteel
Onderzoek der Materie (FOM) and the Nederlandse Organisatie voor
Wetenschappelijk Onderzoek (NWO), the Nanonext consortium and is part of an industrial
partnership program between Philips and FOM. This research has also
received funding from the European Community's Seventh Framework Programme under Grant Agreement no. 214954. G. Lozano thanks NanoNextNL for funding his postdoctoral contract.

\bibliographystyle{unsrt}
\bibliography{bibliography}


\newpage
\begin{figure}
\centerline{\includegraphics[width=100mm]{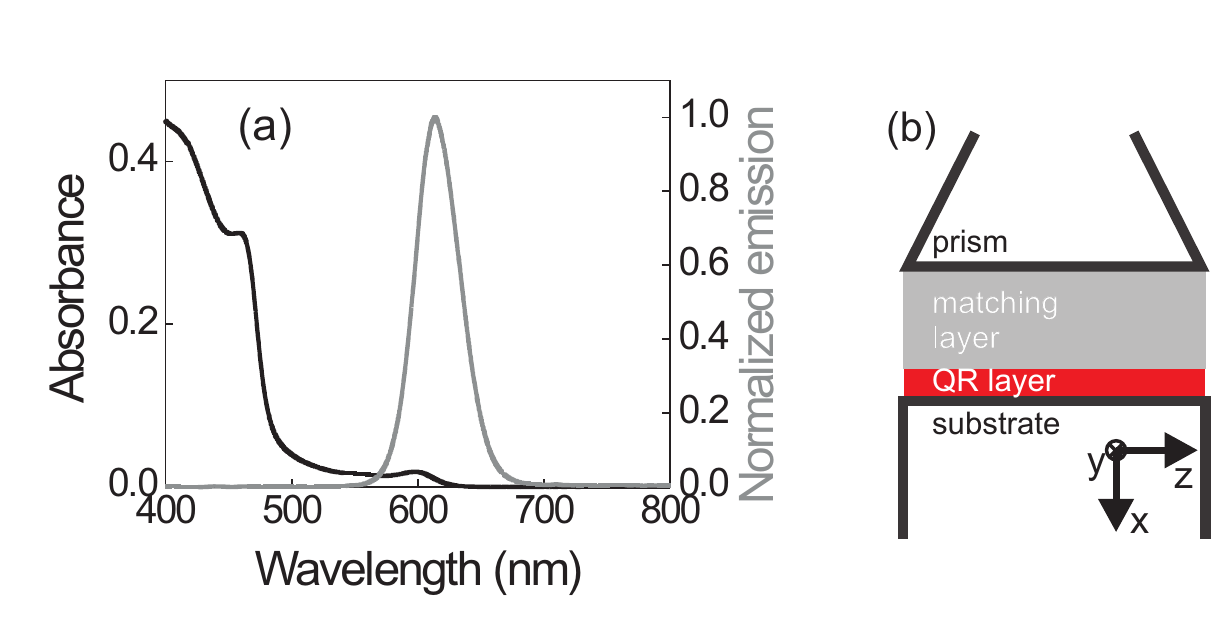}}
\caption{(Color online) (a) Absorbance (black curve) and  normalized emission spectra (gray curve)
of CdSe/CdS quantum dot-in-rods. (b) Schematic representation
of the sample, i.e., F2 glass prism, silica matching layer, QR
layer and quartz substrate.}\label{fig1-eps-converted-to.pdf}
\end{figure}

\newpage
\begin{figure}
\centerline{\includegraphics[width=100mm]{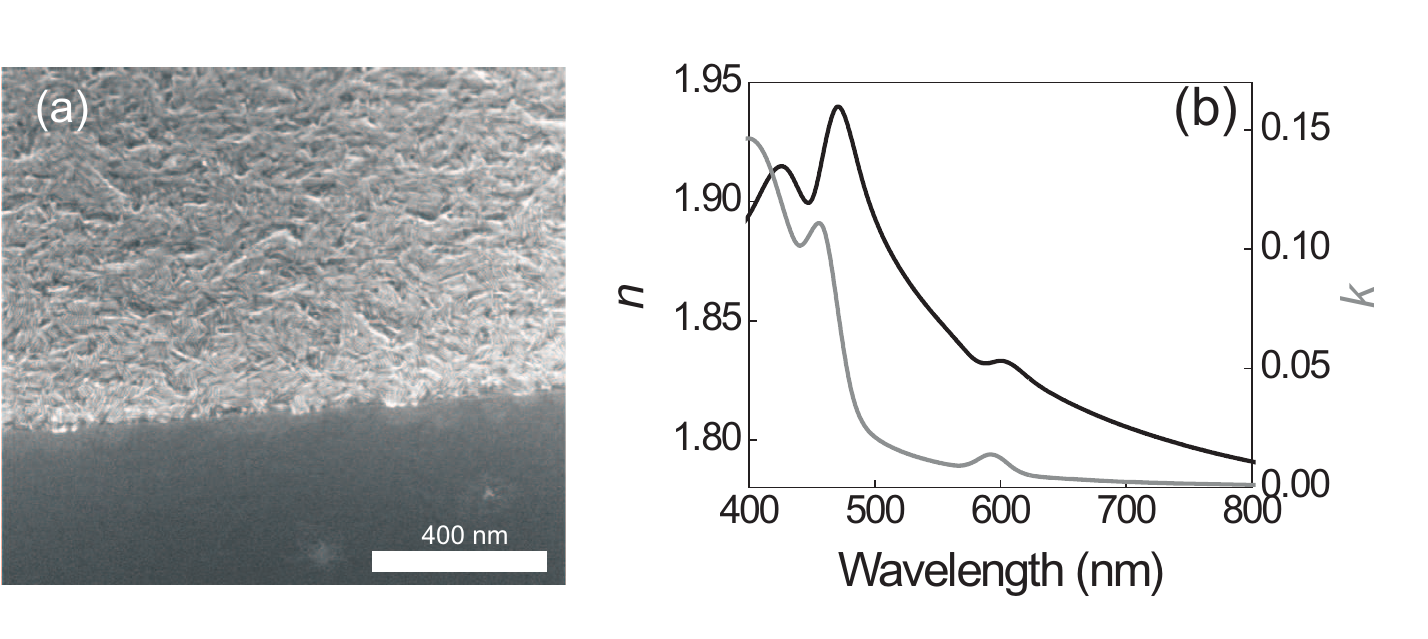}}
\caption{(a) Tilted scanning electron microscope image of the top surface of the
QR layer spin-coated over a quartz substrate. (b) Real (black
curve) and imaginary (gray curve) components of the refractive index
of the QR layer as a function of the wavelength.}\label{fig2-eps-converted-to.pdf}
\end{figure}

\newpage
\begin{figure}
\centerline{\includegraphics[width=160mm]{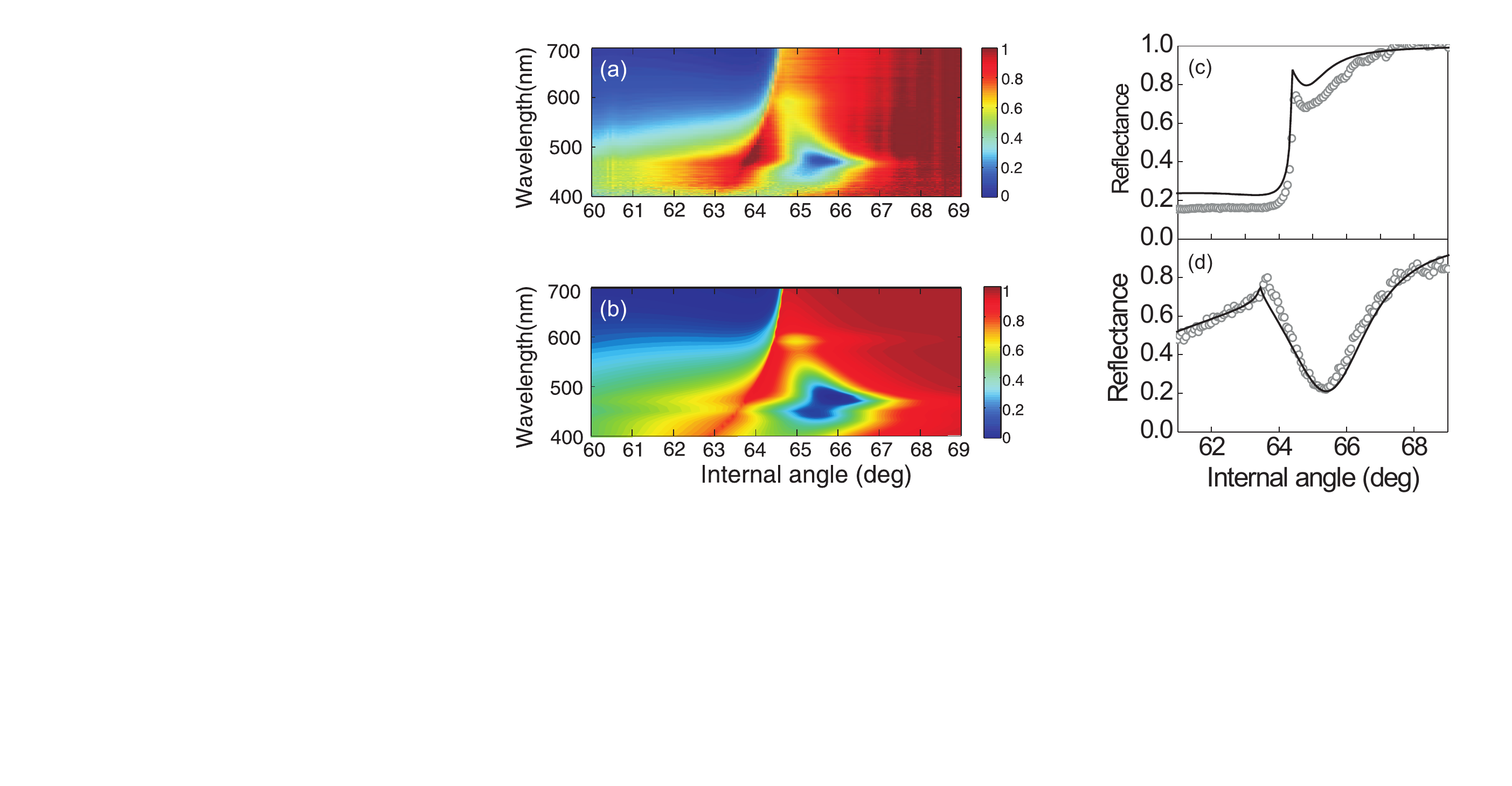}}
\caption{(a) Measured and (b) calculated specular reflectance spectra from the
sample represented in Fig. 1(b) of \textit{s}-polarized light (color scale).
The reflectance is displayed as a function of the wavelength and the
angle of incidence. (c) and (d) are the experimental (open circles)
and calculated (solid lines) specular reflectances at $\lambda$= 611
nm and $\lambda$= 457 nm, respectively.}\label{fig3-eps-converted-to.pdf}
\end{figure}

\newpage
\begin{figure}
\centerline{\includegraphics[width=100mm]{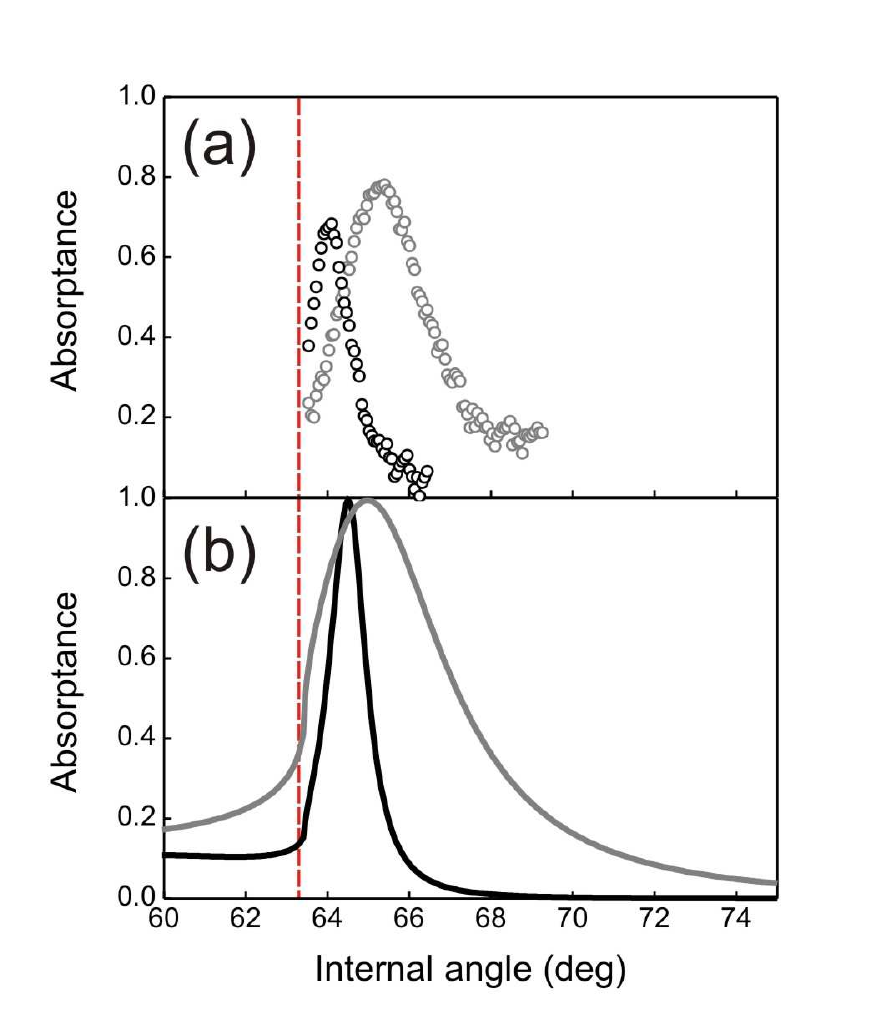}}
\caption{(Color online) (a)
Experimental absorptance defined as 1-\textit{R} of $\lambda$=457 nm \textit{p}-polarized (open black
circles) and \textit{s}-polarized (open gray circles) incident light onto the
sample schematically represented in Fig. 1(b), as a function of the
angle of incidence. (b) Calculated absorptance for \textit{p}-polarized (black
curve) and \textit{s}-polarized (gray curve) incident light with an optimized
thickness of the matching layer [see Fig. 1 (b)] in order to obtain
maximum absorptance. The red dashed line indicates the
critical angle for total internal reflection at the interface
between prism and matching layer.}\label{fig4-eps-converted-to.pdf}
\end{figure}

\newpage
\begin{figure}
\centerline{\includegraphics[width=100mm]{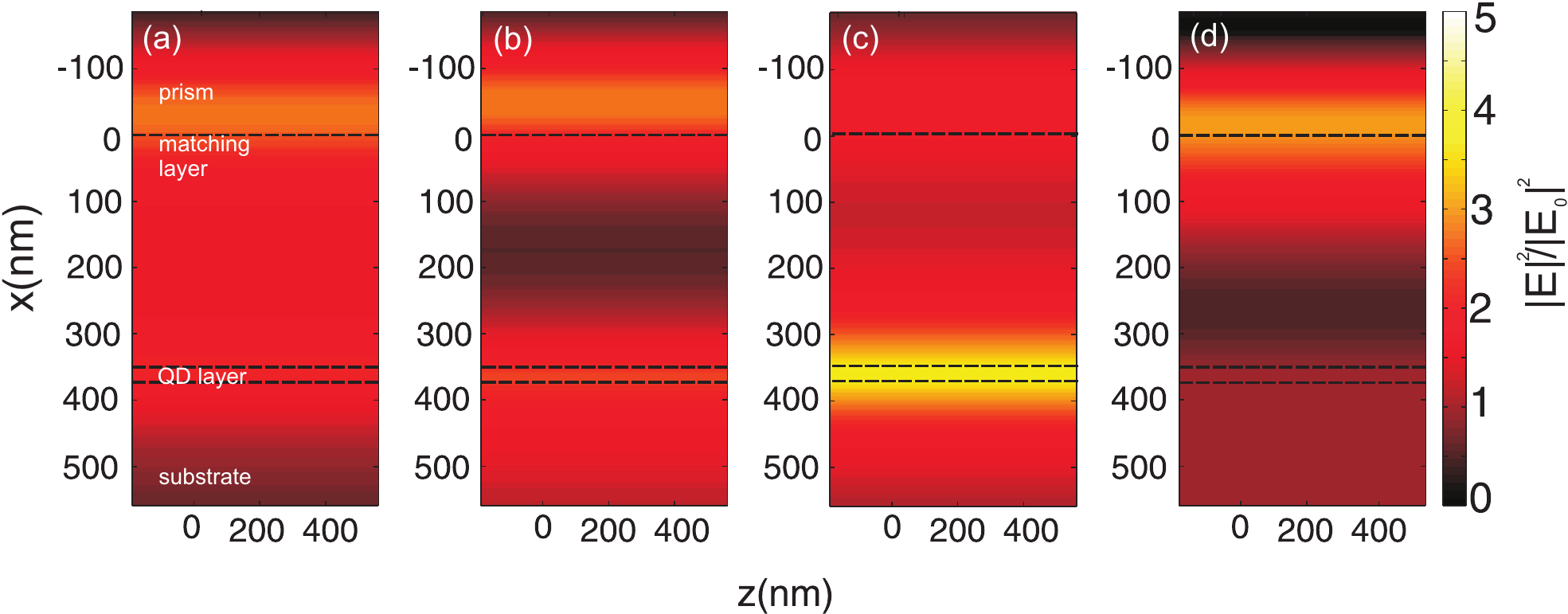}}
\caption{(Color online) Calculated spatial distribution of the electric field
intensity normalized by the incident electric field intensity along
the cross section of the sample schematically represented in Fig.
1(b). In this calculation an incident plane wave of $\lambda=457$ nm
and \textit{s}-polarizion impinges at an angle of incidence of
$\theta=62^\circ$ (a), $64.2^\circ$ (b), $65.4^\circ$ (c) and
$66.5^\circ$ (d). }\label{fig5-eps-converted-to.pdf}
\end{figure}

\newpage
\begin{figure}
\centerline{\includegraphics[width=100mm]{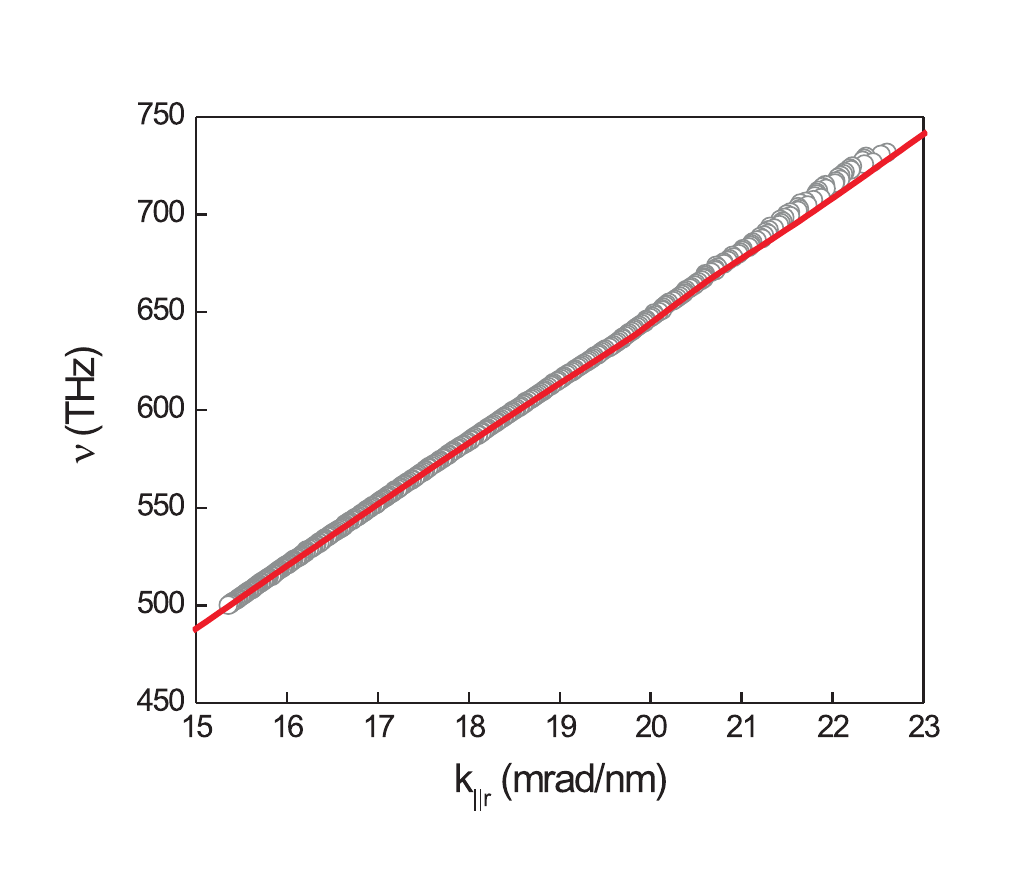}}
\caption{(Color online) Dispersion relation of the ${\rm TE}_0$ mode
in a homogeneous layer with a thickness of 23 nm and a permittivity given in Fig. 2 (b) (red line). Open circles are the experimental data extracted from the
measurements of Fig. 3(a).}\label{fig6-eps-converted-to.pdf}
\end{figure}

\newpage
\begin{figure}
\centerline{\includegraphics[width=100mm]{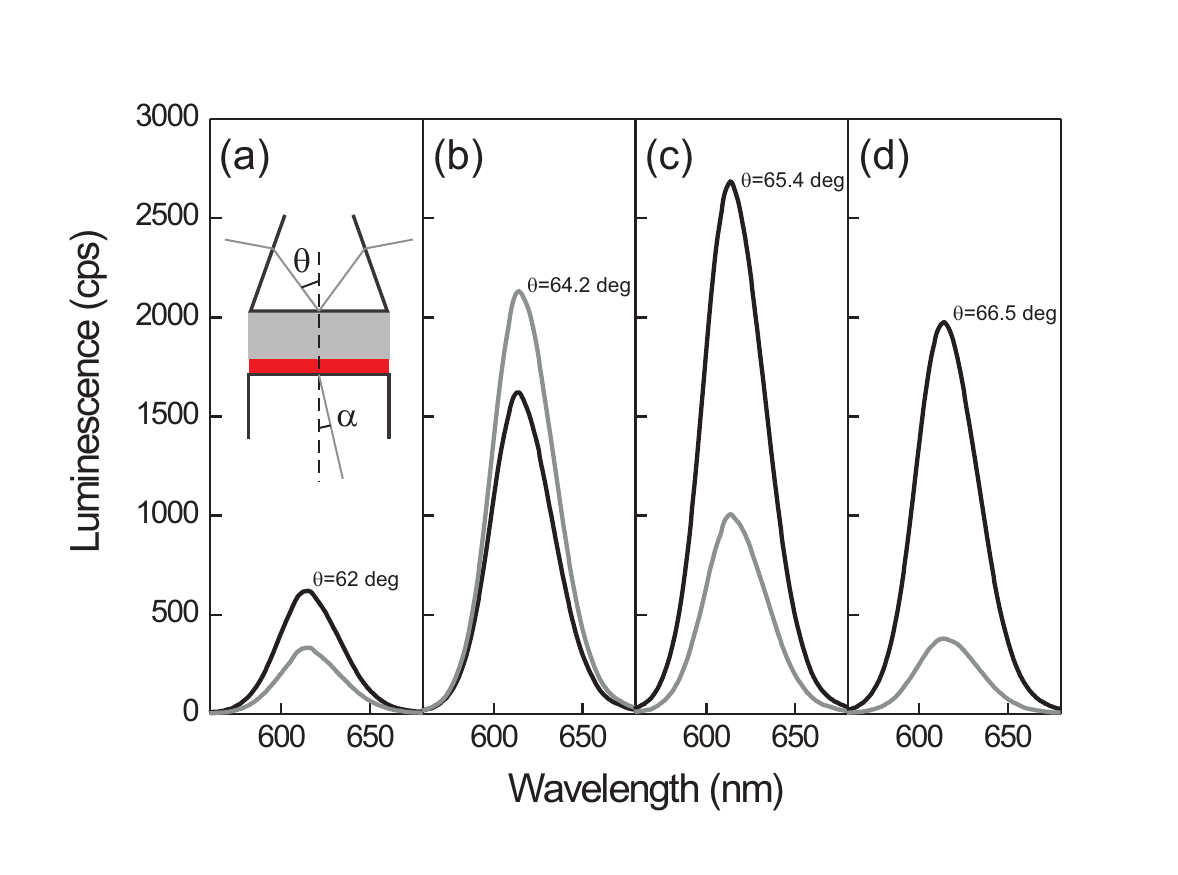}}
\caption{(Color online) Unpolarized photoluminescence spectra measured from the QR
layer schematically represented in Fig. 1(b). The QRs are excited
using \textit{s}-polarized (black curves) and \textit{p}-polarized (gray curves) light
of $\lambda=457$ nm (1.6 mW) incident at (a)
$\theta=62^\circ$, (b) $\theta=64.2^\circ$, (c) $\theta=65.4^\circ$
and (d) $\theta=66.5^\circ$. The collection angle (a) was fixed to
$37^\circ$ in all the experiments.}\label{fig7-eps-converted-to.pdf}
\end{figure}

\newpage
\begin{figure}
\centerline{\includegraphics[width=100mm]{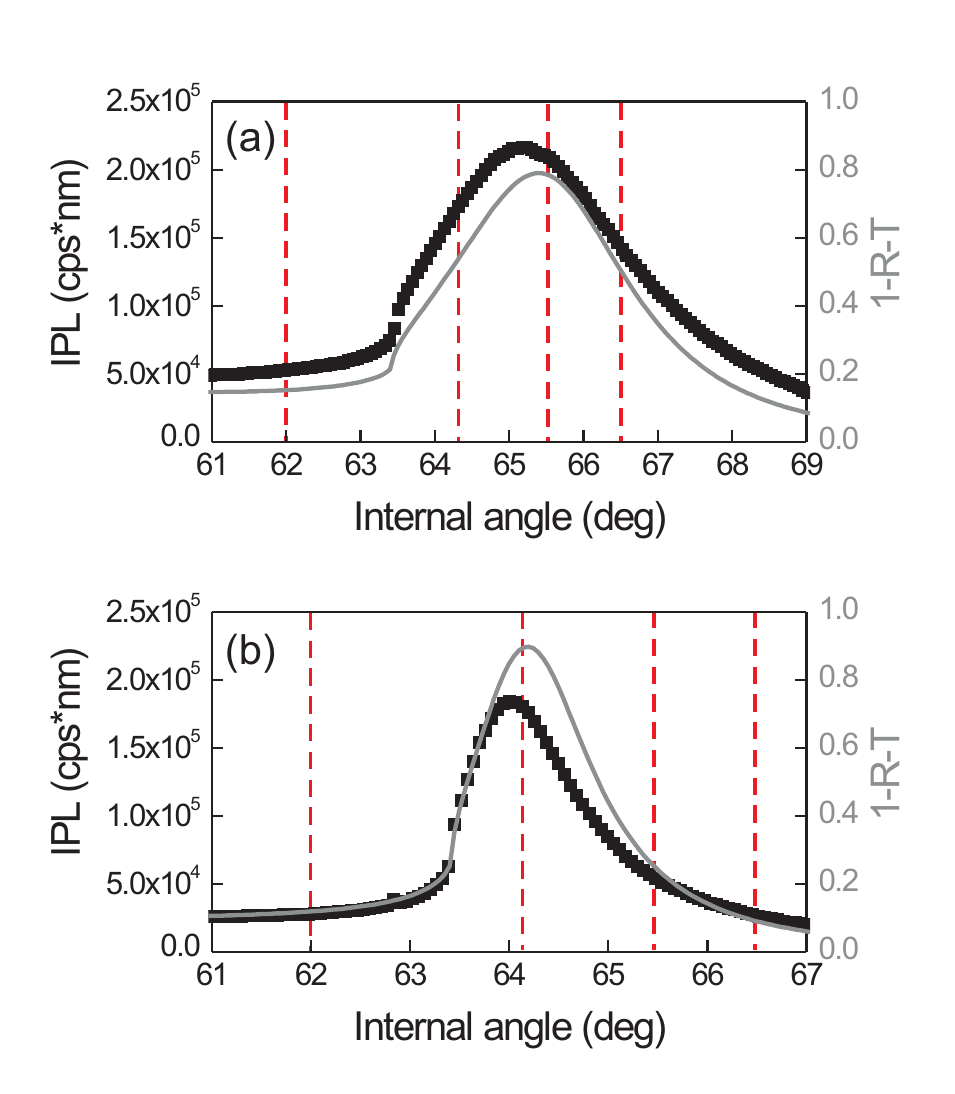}}
\caption{(Color online) Measured photoluminescence intensity (black squares) and
calculated absorptance (gray curves) of the QR layer schematically
represented in Fig. 1(b), versus the angle of incidence of an (a)
\textit{s}-polarized and (b) \textit{p}-polarized beam of $\lambda=457$ nm. The
vertical dashed lines indicate the angle of excitation of the
photoluminescence spectra shown in Fig. 7. }\label{fig8-eps-converted-to.pdf}
\end{figure}

\end{document}